\documentstyle[12pt,aaspp4,psfig]{article}

\newcommand{\etal} 	{{et~al.}}
\newcommand{\kms}	{km~s$^{-1}$}
\newcommand{\hst}	{{\it HST}}
\newcommand{\lya}	{Ly$\alpha$}
\newcommand{\pcm}	{cm$^{-2}$}
\newcommand{\apl}  	{^{<}_{\sim}}

\received{July 8th 1996}
\accepted{July 29th 1996}

\begin{document}

\title{
The Absence of Diffuse Gas around the Dwarf Spheroidal Galaxy Leo I$^1$
}

\author{David V. Bowen$^{2}$, Eline Tolstoy$^{3,4}$, Andrea Ferrara$^{5}$, 
J. Chris Blades$^{3}$, Elias Brinks$^{6,7}$}

\affil{ $\:$}

\affil{$^2$Royal Observatory, Edinburgh, Blackford
Hill, Edinburgh EH9 3HJ, United Kingdom\\email: dvb@roe.ac.uk}

\affil{$^3$Space Telescope Science Institute, 
3700 San Martin Drive, Baltimore, MD 21218\\email: tolstoy,blades@stsci.edu }

\affil{$^5$Osservatorio Astrofisico di Arcetri, Largo E Fermi 5,50125
Firenze, Italy\\email: ferrara@arcetri.astro.it}

\affil{$^6$NRAO, P.O. Box O, 1003 Lopezville Rd., Socorro, NM 87801-0387\\
email: ebrinks@nrao.edu}

\altaffiltext{1}{Based on observations obtained with the NASA/ESA Hubble 
Space Telescope, obtained at STScI, which is operated by the Association of
Universities for Research in Astronomy, Inc., under contract with the National
Aeronautics and Space Administration, NAS5-26555.}
\altaffiltext{4}{Present address: ST-ECF, ESO, Karl-Schwarzschild-Strasse 2,
D-85748, Garching bei M\"{u}nchen, Germany}
\altaffiltext{7}{Currently on leave of absence at the Departmento de
Astronomia, Universidad de Guanajuato, Apdo.\ Postal 144, Guanajuato,
C. P. 36000, Mexico}

\begin{abstract}

We have obtained spectra of 
three QSO/AGNs with the GHRS aboard the {\it Hubble
Space Telescope} to search for absorption from 
low column density gas in the halo of the dwarf
spheroidal (dSph) galaxy Leo~I. The probe sightlines 	    
pass 2.1, 3.7, and 8.1~kpc from the
center of the galaxy, but no C~IV, Si~II, 
or Si~IV absorption is found at the
velocity of Leo~I. The absence of low ionization species
suggests that the column density of
neutral hydrogen which exists within $2-8$~kpc of the galaxy is 
$N$(H~I) $\apl 10^{17}$~\pcm ; assuming that the high ionization lines of
Si~IV and C~IV dominate the ionization fraction of silicon and carbon, 
the limit to the total hydrogen column is $N$(H) $\apl 10^{18}$~\pcm .

Our results demonstrate that there are no dense flows of gas in or out of
Leo~I, and that there is no evidence for tidally disrupted gas which might
have accompanied the galaxy's formation or evolution.  However, our detection
limits are insufficient to rule out the existence of a sphere or shell of
ionized gas around the dSph, with a mass up to that constituting the entire
galaxy. Our models show that dSph galaxies similar to Leo~I are not massive
enough to have halos which can contribute significantly to the metal line
absorption cross-section of QSO absorbers seen at high redshift.

\end{abstract}


\section{Introduction}

That there is still no consensus as to how dwarf galaxies form is demonstrated
by the number of theories which exist to explain their origin. The dwarf
spheroidal (dSph) galaxies, in particular, present a challenge---and indeed a
constraint---to our understanding of galaxy formation and evolution. The
properties of these galaxies, and the theories which are advanced to explain
their origin and evolution, are summarised in detail by Gallagher \& Wyse
(1994). Notably, many of the hypothesised mechanisms imply that the dSph
galaxies we see today could be surrounded by extended gaseous halos. In this
paper we present the results for a search for such a halo around Leo~I,
using UV absorption lines to expose the existence of any low density gas
around the galaxy which cannot be detected in any other way.

How might the formation and evolution of dwarf galaxies influence the
distribution of gas around them?
Dwarf galaxies may collapse and evolve almost
independently from massive galaxies, forming in large numbers at early epochs
from primordial density fluctuations (Ferguson 1994).  Dwarf irregular galaxies
may evolve from dwarf ellipticals by accreting gas cooling from the
intergalactic medium (Silk, Wyse \& Shields 1987).  Alternatively, dwarf
galaxies may act as the basic building blocks of all galaxies, merging at
higher redshift to form the distribution of galaxies we see today.  Dwarfs
seen at the present epoch would then be the few remnants from this earlier
period of galaxy formation.  With more massive galaxies in place,
dwarf spheroidals may form as the result of interactions between galaxies
(Gerola, Carnevali \& Salpeter 1983; Hunsberger, Charlton \& Zaritsky 1996),
or they may evolve from more massive elliptical galaxies suffering substantial
gas loss through SN-driven winds (Vader 1986).

Similarly, dSph galaxies may form and evolve as a result of mass loss from the
cumulative effect of supernovae and stellar winds in more gas rich systems.
Such processes would be highly effective in re-distributing gas away from the
center of the galaxy.  Supernovae may drive gas out of low mass (proto-)
galaxies before most of the initial gas reservoir is converted into stars
(Larson 1974; Saito 1979; Dekel \& Silk 1986, Ferrara \& Tolstoy 1996).
Bursts of star formation lasting $> 10^{8}$ years would then deposit large
amounts of energy into the surrounding interstellar medium, imparting enough
momentum for the (metal enriched) gas to become unbound.  The gas then escapes
the galaxy and mixes with the intergalactic medium (e.g.  De Young \&
Gallagher 1990; De Young \& Heckman 1994).  Though these processes may be at
work in all types of dwarf galaxies, extensive mass loss may weaken the
potential well of the lowest mass galaxies, producing relatively round, low
surface brightness and low metallicity remnants similar to the dwarf
spheroidals seen in the Local Group (Saito 1979), such as Leo~I.

If the galaxy mass is high enough, it is possible that the flow breaking
out of the main body of the galaxy will remain bound to it.
This `dwarf galactic fountain' could then eventually fall back 
to the center of the
galaxy causing subsequent bursts of star formation.  Recent analysis of
color-magnitude diagrams of the stellar populations of Carina and Leo~I
suggest that they have undergone more than one discrete burst of star
formation (Smecker-Hane~\etal\ 1994; Lee~\etal\ 1993), which would support
this theory.  Clearly, such evolution should produce multiple shells of gas
spread over many kpc which ought to be detectable through UV absorption lines.

Thus, discriminating between different theories of dSph galaxy evolution has
important implications for our general understanding of galaxy formation.
Importantly, most of these scenarios suggest that interstellar gas will be
substantially disrupted, leading to its redistribution away from the stellar
population.  Such gas is likely to be of low column density, however, and
susceptible to ionization by the intergalactic UV background, making it
difficult to observe.  Gas which is shocked due to its expulsion from a galaxy
will also remain highly ionized.  The only way to detect the gas is to search
for the weak UV absorption lines it produces in the spectra of background
sources.  In this paper, we report on a search for C~IV, Si~IV \& Si~II
absorption toward 3 QSO/AGNs which lie 2.1, 3.7 and 8.1 kpc from the center of
Leo~I.  Leo~I is particularly interesting because it contains a 
relatively young stellar population, with an age 
measured to be 1.5~Gyr (Caputo, Castellani, Degl'Innocenti 1996),
3~Gyr (Lee~\etal~1993), and $5\pm 2$~Gyr (Demers~\etal~1994). These first two
estimates would place the stellar population's formation at redshifts of $z <
0.5$ for any value of $h$ and $\Omega_b$ in the usually accepted range
$h=0.5-1.0$ and $\Omega_b < 0.2$ ($ \Omega_b$ is the baryonic
density parameter, and $H_0 = 100 h $~\kms~Mpc$^{-1}$, where $H_0$ is the
Hubble constant). Around Leo~I,
therefore, there may still exist some evidence of an extended gas envelope
which formed during the galaxy's evolution.

In \S2 of this paper we outline the observations made with \hst\ and present
the spectra obtained of the three probes, which show no absorption lines from
Leo~I. In \S3 we list the equivalent width limits obtained and their
conversion to column densities. We calculate a limit to the total gas
column density in \S4.1, along with estimates of the gas volume density and
mass within the inner radius of the halo, assuming two different models for
the distribution of gas around the galaxy. Finally, in \S4.2, we
briefly explore the implications of our results for the hypotheses of galaxy
formation and evolution discussed above.

\section{Observations and Data Reduction}

Observations of Q1004+1303, Q1008+1319, and Q0957+1317 were made 17-May-1995,
03-June-1995, and 29-May-1995, respectively, with the Large Science Aperture
(LSA) of the GHRS and the G140L grating centered at 1416~\AA .  Q0957+1317 is
actually NGC~3080, a bright Seyfert~1 AGN galaxy also identified as Mrk~1243.
The $V$-band magnitudes, and the redshifts of the probes, are given in Table
1, along with the observed flux, $F_\lambda$ at 1400~\AA , the separation
between the QSO/AGN and the center of Leo~I on the plane of the sky in
arcmins, $\rho$, and the corresponding separation, $s$, at Leo~I's distance
from us of 210~kpc. Also listed are the separations in terms of the tidal
radius, $r_c$. Q1004+1303 and Q1008+1319 are the closest objects to Leo~I on
the plane of the sky, within $\simeq$ 1 degree. We reproduce in
Fig.~\ref{fig:poss} a region of the sky around the galaxy taken with the UK
Schmidt Telescope showing the positions of these two objects relative to
Leo~I. 
{\tt This plate is unavailable with this Eprint}

Exposure times were the same for each object, 1.63 hrs, (3 orbits).  The data 
were taken with an FP-SPLIT of two, and quarter-stepped, giving a 
dispersion of 0.14~\AA\ pix$^{-1}$.  Time spent measuring the background 
was 11\% of the total exposure time.
The spectra were calibrated using the standard pipeline software.  Data 
sets taken at different carousel positions were coadded, wavelength 
calibrated, and resampled to a linear dispersion using the calibration 
wavelength files.  The offsets in wavelength between the two FP-split data 
were then calculated with the STSDAS IRAF routine {\tt poffsets} and a 
correction applied.  For the Q1008+1319 spectra, the peak of the 
cross-correlation function was poorly determined because of the low 
signal-to-noise of the data; however, for these, and for the two other 
data sets, careful comparison was made between the wavelengths of features 
in the FP-SPLIT data to ensure that the co-addition of the two halves of 
the data was correct.  Error arrays were constructed in the same manner 
using the pipeline error files.

To obtain an exact zero point for the wavelength calibration of the final
coadded data, we compared the velocity of O~I$\lambda 1302$ and Si~II~$\lambda
1526$ absorption lines in the QSO/AGN spectra from the Milky Way, with the
velocity of H~I emission along the line of sight to Leo~I taken from the
Leiden/Dwingeloo 21~cm H~I survey (Hartmann 1994).  For Q1004+1303 and
Q0957+1317, shifts of $-0.4$ and $-0.2$~\AA\ (0.7 and 0.4 diodes) were
required, respectively.
For Q1008+1319, the data were of too low a signal-to-noise to 
accurately measure the centres of the O~I$\lambda 1302$ and Si~II~$\lambda 
1526$ absorption lines, but their positions are consistent with that 
expected from the 21~cm measurement, and no correction was applied.

Plots of the normalised spectra around the position of the Si~IV lines 
are shown in Figure~\ref{fig:spec1}, while portions of the spectra around 
C~IV are shown in Figure~\ref{fig:spec2}. The figures mark the rest 
wavelengths of the absorption expected from our own Galaxy, as well as 
the wavelengths of any absorption arising from Leo~I. In this case we 
take the velocity of Leo~I to be 285~\kms\ (Zaritsky~\etal\ 1989). 

\section{Results}

As Figures~\ref{fig:spec1} and~\ref{fig:spec2} show, there is no evidence for
Si~IV or C~IV absorption from Leo~I.  2$\sigma$ equivalent
width limits, $2\sigma$($W$), to the absorption are given in Table~2a.
$\sigma$($W$) is calculated from $\sigma$($W$)$^2\:=\:\delta \lambda^2 \sum{N}
\sigma_i^2$ where $\sigma_i$ is the error in the measurement of the flux at
the $i$th pixel (measured from the calibrated error arrays), $N$ is the number
of pixels the line is measured over, and $\delta\lambda$ is the dispersion.
The Line Spread Function for the GHRS taken after the installation of {\tt
COSTAR} is approximately Gaussian with a width of 1.4 diodes FWHM, or for the
data discussed herein, 5.6 pixels.  We have therefore taken $N$ to be 11.
Table~2b lists the equivalent widths, $W$, of the Milky Way absorption lines.
Figure~\ref{fig:spec1} shows that the Si~IV$\lambda 1392$ line seen in the
spectrum of Q1008+1319 is extremely strong and resolved, considerably stronger
than the absorption seen towards the other two lines of sight.  Yet the
corresponding Si~IV$\lambda 1402$ line is absent.  Either the Galactic
Si~IV$\lambda 1392$ has an equivalent width several $\sigma(W)$ from its
correct value, or it is actually blended with a stronger higher-redshift
absorption line (possibly \lya\ at $z = 0.146$) and does not represent Si~IV
absorption from our own Galaxy.

To calculate limits to the column densities of the gas, we assume that any gas
which has not been detected would give rise to absorption lines with
equivalents widths derived from the linear part of the curve of growth.  For
the limits listed in Table~2a, lines are independent of the doppler parameter,
$b$, for $b \gtrsim 15-20$~\kms .  
For sightlines through our own Galaxy toward
extragalactic sources, or in high redshift QSO absorption line systems, $b$
values of $10-20$~\kms\ are measured for most C~IV and Si~IV lines
(e.g. Savage, Sembach \& Lu 1995, and refs therein; 
Fan \& Tytler 1994; Lu~et~al.\ 1994).  These
lines are observed at $10-20$~\kms\ resolution, and may be comprised of
several components.  The few observations of C~IV and Si~IV absorption lines
taken at the highest resolution ($\sim 3.5$~\kms , with the echelle of the
GHRS), where individual components might be observed, are along sightlines
through the Milky Way, for which values of $b$ between $5- 12$~\kms\ for Si~IV
and $10-27$~\kms\ for C~IV are found (Savage, Sembach \& Cardelli 1994;
Sembach, Savage \& Jenkins 1994).  
These lines may themselves be comprised of components which are not resolved
even at this high resolution, but 
since single, isolated lines with small $b$ values are rarely seen, 
a limit of $b \gtrsim 15-20$~\kms\ is probably adequate to characterise any
absorption close to our equivalent width limit.

Table~2a lists the equivalent width limits to the Si~IV and C~IV absorption
lines.  No useful limits can be obtained for the absorption toward Q1008+1319
due to the low signal-to-noise of the data, but for the remaining two
sightlines, we can derive column density limits for the following ions:

{\it Si~IV:} Toward Q1004+1303 and Q0957+1317 the limit to the column density
of $N$(Si~IV) is almost the same, $\log N$(Si~IV)$ < 12.9$ and $<13.1$.  

{\it C~IV:} The limit to $N$(C~IV) toward Q1004+1303 is $\log N$(C~IV)$ <
13.4$; 
towards Q0957+1317, $\log N$(C~IV) $< 13.8$.

{\it Si~II:} We can also derive a limit to the Si~II column density from the
lack of the Si~II$\lambda 1526$ line, since the limit 
to the equivalent width is
the same as that for the C~IV line: 
towards Q1004+1303, $\log N$(Si~II)$ < 13.4$, while for 
Q0957+1317, $\log N$(Si~II)$ < 13.8$.

{\it Mg~II:} The GHRS spectrum taken by Bowen, Blades \& Pettini (1995;
hereafter BBP) allows us to place a tight
constraint on the Mg~II column density toward Q1004+1303. BBP set an
equivalent width of 40m\AA , which corresponds to $\log N$(Mg~II)$ < 12.0$.

The sightline toward Q1004+1303 provides the lowest column density limits at
the closest impact parameter, as well as an additional measurement of
$N$(Mg~II) from BBP. We therefore collate and summarise these limits in Table
3, although as can be seen from the results above, the limits to the column
densities toward Q0957+1317 are similar (although no search for Mg~II
absorption has been made along this sightline).

\section{Discussion}

\subsection{Limits to the gas mass and gas density around Leo~I}

To understand whether the lack of absorption in the halo of Leo~I is
significant, we need to estimate the limit to the total column density of gas
along the QSO/AGN lines of sight.  To convert to total column densities of
carbon and silicon (summed over all ionization stages) we need to know the
ionization state of the gas. That is, we need to know whether C~IV or Si~IV
was not detected because the majority of the gas lies in a different
ionization stage.

The absence of Si~II, and particularly Mg~II absorption towards Q1004+1303 to
good column density limits, suggests that any gas which is undetected is
probably optically thin at the Lyman limit, so that the H~I column density,
$N$(H~I), is less than $ 2\times 10^{17}$~cm$^{-2}$.  For example, simple
ionization models show that Mg~II disappears rapidly as H~I becomes optically
thin at the Lyman limit, (e.g.  Bergeron \& Stasi\'{n}ska 1986; Steidel \&
Sargent 1992), falling below $10^{12}$~cm$^{-2}$---the limit we measure
towards Q1004+1303---as $N$(H~I) drops below $2\times 10^{17}$~cm$^{-2}$.
Also, the lack of any detectable H~I around Leo~I from 21~cm measurements
(Knapp~et~al.~1978) to a limit of $M_{\rm{HI}} < 7.2\times10^3 M_{\odot}$ also
strongly suggests that there is no optically thick gas anywhere near the 
lines of sight.

Thus, to calculate limits to the total column densities of carbon and silicon,
$N$(C) and $N$(Si), along the lines of sight, we assume that undetected gas is
highly ionized, and that a significant fraction of it is in the form of C~IV
or Si~IV.  This need not be so: models of the fractional ionization of
different metals photoionized by a UV background by Donahue \& Shull (1991)
show that C~IV and Si~IV rarely dominate the ionization fractions in the
gas. However, they remain significant over several dex in the ionization
parameter, $U = n_\gamma / n_H$, where $n_\gamma$ and $n_H$ are the ionizing
photon and hydrogen densities respectively.  Further, if the gas was
collisionally ionized alone, Si~IV would cease 
to contribute more than 30\% of the
ionization fraction at temperatures of $\log T \gtrsim 5.5$ (Shull \& Van
Steenberg 1982).  
If a significant fraction
of the gas is {\it not} in the form of C~IV and Si~IV, the
implication is that gas around Leo~I is extremely hot and highly ionized, and
that $N$(H) derived below is under-estimated.

We also note that Donahue \& Shull (1991) conclude that the resulting limit on
$U$ for the narrow metal line systems observed at redshifts of $z>2$ is $-3.1
\leq U \leq -2.1$.  At these redshifts the ionizing flux---and hence $U$---is
expected to be larger than the present day value.  Yet C~IV and Si~IV only
fail to contribute significantly to the ionization fraction of C and Si for
$\log U > -1$.  Hence if any (undetected) gas around Leo~I was similar to that
observed in higher redshift QSO absorption line systems, the possibility of
ionization stages higher than C~IV and Si~IV contributing more significantly
to the total amount of gas appears to be ruled out.

The total hydrogen column density, $N$(H), is related to the metal line column 
densities by

\begin{equation}
\log N({\rm{H}}) = \log N({\rm{X}}) - D_X - A_X ,
\end{equation}

where $\log N$(X) is the column density of a particular element X, $D_X$ is
the gas phase abundance of element X compared to its solar value, defined as
$\log N\rm{(X/H)} - \log N\rm{(X/H)}_\odot$, or, equivalently, $\log
N{\rm{(X/H)}} - A_X$, with $A_X = \log N\rm{(X/H)}_\odot$ the solar abundance
of X.  So if C~IV and Si~IV contribute significantly to the ionization
fractions of carbon and silicon, $N$(C)$\approx N$(C~IV), and $N$(Si)$\approx
N$(Si~IV), we can calculate a limit to $\log N$(H).  We take ($A_X + 12.00$)
to be 8.65 and 7.57 for carbon and silicon, respectively (Morton, York, \&
Jenkins 1988). $D_X$ is not known for interstellar gas in or around Leo~I; gas
around the galaxy is unlikely to be more metal rich than the stellar
population, but again, the metallicity of the stars in not well determined.
Values of [Fe/H] = $-$1.6, (Demers, Irwin \& Gambu 1994) , $-$2.0 (Lee~\etal\
1993) and $-$(0.7$-$0.3) (Reid \& Mould 1991) have been measured; for our
estimate of $\log N$(H), we adopt a value of 1/10 solar, $D_X = -1.0$.

Values of $\log N$(H) for Q1004+1303 are given in Table 3, and are $<18.5$
derived from the limit to the Si~IV absorption, and $<18.0$ from C~IV.  As
noted in \S3, the values for Q0957+1317 are similar.  The table also includes
the values which would be derived from Mg and Si assuming Mg~II and Si~II
dominated their respective ionization stages, for comparison.  (We take $A_X +
12.00 = 7.60$ for magnesium).  These values would give $N$(H) if our
assumption that the gas was optically thin was incorrect, and lower ionization
species dominated.  We note that the absence of Mg~II absorption lines would
give $\log N$(H) $ < 17.4$.

Although we can obtain little information on the column densities toward
Q1008+1319, 3.7~kpc from the center of Leo~I, the two brighter
objects allow us to quantify the column density of gas 2.1 and 8.1~kpc from
the center.  We
conclude that for Leo~I, the lack of low ionization absorption lines suggest
$\log N$(H~I)$ < 17$, and that the total hydrogen column is
$\log N$(H)$ < 18$, at separations of $2-8$~kpc from
the center of the galaxy. This limit to $N$(H) 
is too small if the gas is
hotter and more highly ionized, or the gas phase abundance, $D_X$, is
$< -1$.

To calculate a limit to the mean density of hydrogen around Leo~I, $\rho$, and
the mass of hydrogen, $M_{\rm{H}}$, we consider two possible geometries for the
distribution of any gas which may remain around the galaxy. We consider (a)
that the gas resides in a spherical halo of radius $R_s$, or (b) that the gas
resides in a shell of thickness $\ell$ and outer radius $R_s$.  Physically,
the two models are important because they could plausibly arise from outflows
of gas as a result of processes within the ISM of the galaxy.
Fig.~\ref{fig:spec3} shows the limits to $\rho$ and $M_{\rm{H}}$, for $\log
N$(H)$ = 18$ and a QSO/AGN-galaxy separation of $2$~kpc (although the results
are practically independent on this latter value) as a function of the assumed
outer radius of the gaseous halo, $R_s$. Values of $\rho$ and $M_{\rm{H}}$ can
be read off the figure from the lines marked $\rho$ and $M$ for any adopted
value of $R_s$.

In the case where the gas resides in a shell, it is necessary to make an extra
assumption about its thickness.  The cooling length behind the radiative shock
leading to the shell formation is $ \ell \sim v_s t_{cool}$, where $v_s$ is
the shock velocity, and $t_{cool}=kT/\rho_{\rm{IGM}}\Lambda(T)$ is the cooling
time for the shocked gas at temperature $T$; $\rho_{\rm{IGM}}$ is the density of
the ambient (i.e.  intergalactic) medium (Giroux \& Shapiro 1996) which we
take to be $8.6\times 10^{-6}\Omega_b (1+z)^{3} h^2$~cm$^{-3}$ or $7.7\times
10^{-8}$~cm$^{-3}$ for $\Omega_b = 0.009$ and $h=1$ [this is 
the lower limit to $\Omega_b h^2$ given by Copi, Schramm \& Turner (1995), $
0.009 \leq \Omega_b h^2 \leq 0.02$; adopting the upper limit does not change
our results]. 
$\Lambda(T)$ is the cooling rate.  If $v_s\simeq v_e$, where $v_e=15$~\kms\ is
the escape velocity from the galaxy, then $\ell = 4.4$~kpc.

Fig.~\ref{fig:spec3} shows that for $\log N$(H)$ = 18$ the upper limit to
$\rho$ is $\sim 0.7- 5 \times 10^{-4}$~cm$^{-3}$ for both models, for $R_s =
5-50$~kpc, while the upper limits to $M_{\rm{H}}$ reach, for example, $6\times
10^8 -1\times 10^9 M_\odot$ for the spherical and shell case respectively,
for $R_s=50$~kpc.

\subsection{Has the gas gone?}

Fig.~\ref{fig:spec3} shows that the total mass of gas around Leo~I is not well
constrained by our observations since a shell or sphere of gas could exist
over a wide range of radii ($R_s$). In fact, the total mass of Leo~I is known
from derivations of the global and central $M/L$ ratios by Irwin \&
Hatzidimitriou (1995), who found $M/L \simeq 1$, and therefore $M\simeq
3\times 10^6 M_{\odot}$ for $L\sim 3.4\times 10^6 L_\odot$. From
Fig.~\ref{fig:spec3}, it can be seen that this much mass could only give rise
to a column density of $N$(H)$ =10^{18}$~\pcm\ if $R_s \simeq 7$~kpc and $\rho
\simeq 10^{-5}$~cm$^{-3}$ (there is no solution for a shell since its
thickness, $\ell$, is comparable to $R_s$). With $M_{\rm{HI}} < 7\times10^{3}
M_{\odot}$ (Knapp~\etal\ 1978), this sphere would be highly ionized and would
account for all the observed dynamical mass. For $N$(H) $ << 10^{18}$~\pcm ,
the same---or less---gas mass can be distributed over larger spheres (or
shells).  For example, at $\log N$(H)$ = 15$, a strong constraint on $R_s$
exists because $\rho$ is comparable to $\rho_{\rm{IGM}}$. Since $\rho >
\rho_{\rm{IGM}}$ for a shell or sphere to exist, it is possible to show that
$R_s$ must be less than 5~kpc and that the mass of gas would be $10^4 M_\odot$
and $10^3 M_\odot$ for a shell and halo, respectively. Unfortunately, the
metal absorption line column densities required to obtain these limits to
$N$(H) are very low. For example, $\log N$(C~IV) would have to be $\sim 11$
for gas with 1/10 solar metallicity to reach $\log N$(H)$ = 15$, a column
density unattainable with current instrumentation. A more suitable probe of
low column density H~I would be the \lya\ line, since the transition is
sensitive to H~I column densities several dex less than the metal absorption
lines. Unfortunately, at the velocity of Leo~I, \lya\ absorption would be lost
in the strong absorption from the Milky Way.

The idea that dwarf galaxies may be responsible for both metal-line and \lya\
QSO absorption systems at high redshift has been widely discussed (York~\etal\
1986; Tyson 1988; Impey \& Bothun 1989; Rauch~\etal\ 1996).  Our models show
that low luminosity dSph galaxies like Leo~I are simply not massive enough to
have halos which can be detected from metal absorption lines, even {\it if}
all their mass resides in an ionized halo.  This does not mean
that more massive/luminous dwarfs, including gas rich dwarfs, do not give rise
to absorption lines at high redshift. Nor does it imply that dSphs could not
give rise to \lya\ absorption lines.
It does suggest, however, that the
population of dSphs, which can be so prevalent in environments like the Virgo
Cluster (Sandage~\etal\ 1985), contribute little to the absorption
cross-section of metal absorption lines such as
C~IV, Si~IV, Mg~II, etc.

Despite the fact that we cannot rule out the presence of diffuse, ionized
shells or spheres around Leo~I, we note that the lack of absorption
could also be because gas has been removed via dynamical processes.  The
absence of high ionization lines is consistent with the conclusion that there
are no inflows or outflows of dense gas intercepting the QSO lines-of-sight,
as might be expected, for example, from concentrated galactic fountains or
dense inflows destined to re-ignite star formation.  If the galaxy underwent a
transient period of intense star formation in which most of the gas was
ejected (Larson 1974; Saito 1979; Dekel \& Silk 1986), the gas could have
merged with the IGM for it now to be undetectable.  Assuming that when
blow-out occurred, the shell moved to an escape velocity of $v_e$ after a
short initial transient, the merging time would be $t_m \sim R_s/v_e$, where
$R_s$ is the size of the shell when $\rho = \rho_{\rm{IGM}}$.  For $R_s =
5-50$~kpc and $v_e = 15$~\kms , $t_m \sim 3\times 10 ^{8-9}$~yr. This is less
than or comparable to the age of the stellar population measured for Leo~I
(see \S1) so it is possible that gas has been removed this way.

Gas which has existed around Leo~I may have been stripped via interactions
with neighbouring galaxies.  Indeed, similar hypotheses have been suggested to
account for the {\it origin} of dSph galaxies (e.g. Gerola, Carnevali \&
Salpeter 1983).  It is impossible to generalise about the ability of
interstellar gas to survive such encounters, its physical state, or its
distribution around the parent dSph.  In more massive galaxies, however,
interactions occur such that tidal debris remains optically thick at the Lyman
limit, with column densities high enough to be detected at 21~cm.  Indeed,
such debris offer some of the best material in which to cause absorption lines
in nearby galaxies (BBP; Bowen et al.~1994; Carilli, van Gorkom \& Stocke
1989).  If Leo~I was formed from a more massive, gas rich galaxy, one
might expect the remains of the stripped gas to still be detectable. The lack
of absorption suggests that any stripping which has occurred could not
have taken place recently. 

\section{Summary}

We have searched for absorption lines of C~IV, Si~II, and Si~IV arising in gas
around Leo~I, towards 3 QSOs who lines of sight pass within $\simeq2-8$~kpc of
the galaxy. We have found no absorption, and conclude that at these distances
the column density of neutral hydrogen is $\log
N$(H~I) $ \apl 17$, while the total hydrogen column density is $\log N$(H)
$\apl 18$, assuming the gas has 1/10 solar metallicity and that most of the
gas is in an ionization state whereby C~IV and Si~IV dominate the ionization
fractions.  Our results are consistent with the conclusion that there are no
dense flows of gas in or out of the galaxy, and there is no evidence for
tidally disrupted gas which might have accompanied Leo~I's formation or
evolution. We cannot rule out the possibility, however, of a sphere or shell
of ionized gas around the dSph, with a mass as high as the entire galaxy's
dynamical mass. The fact that our detection limits are insufficient to reveal
such a gaseous halo demonstrates that dSph galaxies similar to Leo~I are not
massive enough to have halos which can contribute significantly to the metal
line absorption cross-section of QSO absorbers seen at high redshift.

\bigskip

It is a pleasure to thank Dap Hartmann for providing H~I spectra from the
Dwingeloo/Leiden 21~cm survey to help calibrate the GHRS data, Jason Cowan
in the ROE photolabs for the reproduction of the field around Leo~I, and, in
particular, Don
Garnett for an important reading of the paper.
A.F. also wishes to thank George Field for enlightening discussions.  The work
described in this paper was funded from grants GO-5451 and GO-3524 .



\clearpage
\begin{deluxetable}{cccccrcc}
\tablenum{1}
\tablewidth{0pc}
\tablecaption{PROBES BACKGROUND TO LEO~I}
\tablehead{
\multicolumn{5}{c}{$\:$ } 
& \multicolumn{3}{c}{Probe-galaxy separations\tablenotemark{a}}\\
\cline{6-8}
\colhead{}  &\colhead{}  &\colhead{}  &\colhead{}  &\colhead{}  &
\colhead{$\rho$}  &\colhead{$s$\tablenotemark{c}}  &\colhead{}  \\
\colhead{QSO/AGN probe} & \colhead{Alias} & \colhead{$V$} 
& \colhead{$F_\lambda$\tablenotemark{b}}
& \colhead{$z_{\rm{em}}$}  &\colhead{($'$)}  &\colhead{(kpc)}  
&\colhead{$\rho/r_c$}  
}
\startdata
Q1004+1303 & 4C+13.41	& 15.2	& 1.0	& 0.240	& 34.0	& 2.1	& 0.6 \nl
Q1008+1319 & \nodata 	& 16.3	& 0.1	& 1.287	& 60.7	& 3.7	& 1.1 \nl
Q0957+1317 & NCG 3080	& 15.0 	& 0.4	& 0.035	& 132.4	& 8.1	& 2.5 \nl
\tablenotetext{a}{Assuming the center of Leo~I is at $\alpha=$10:08:27.39,
$\delta =$ 12:18:27 (J2000.0)}
\tablenotetext{b}{Flux at 1400~\AA , in units of $10^{-14}$ ergs cm$^{-2}$
s$^{-1}$ \AA $^{-1}$}
\tablenotetext{c}{Assuming a distance to Leo~I of 210~kpc (Demers, Irwin \&
Gambu 1994)}
\enddata
\end{deluxetable}

\clearpage
\begin{deluxetable}{ccccc}
\tablenum{2a}
\tablewidth{0pc}
\tablecaption{EQUIVALENT WIDTHS OF LINES 
IN THE HALO OF LEO~I\tablenotemark{a}}
\tablehead{
\colhead{} 
& \colhead{Si~IV$\lambda1392$} 
& \colhead{Si~IV$\lambda 1402$} 
& \colhead{C~IV$\lambda 1548$} 
& \colhead{C~IV$\lambda 1550$} \nl
\colhead{QSO/AGN probe}
& \colhead{(\AA )}
& \colhead{(\AA )}
& \colhead{(\AA )}
& \colhead{(\AA )} 
}
\startdata
Q1004+1303 
& $<0.07$ 
& $<0.07$ 
& $<0.10$ 
& $< 0.12$ \nl
Q0957+1317
& $<0.11 $
& $<0.11 $
& $<0.22 $
& $<0.24 $ \nl
Q1008+1319 
& $<0.37 $
& $<0.42 $
& $<1.23 $
& $<1.61 $ \nl
\tablenotetext{a}{All limits are $2\sigma(W)$}
\enddata
\end{deluxetable}

\begin{deluxetable}{ccccc}
\tablenum{2b}
\tablewidth{0pc}
\tablecaption{EQUIVALENT WIDTHS OF LINES 
IN THE MILKY WAY HALO\tablenotemark{a}}
\tablehead{
\colhead{} 
& \colhead{Si~IV$\lambda 1392$} 
& \colhead{Si~IV$\lambda 1402$} 
& \colhead{C~IV$\lambda 1548$} 
& \colhead{C~IV$\lambda 1550$} \nl
\colhead{QSO/AGN probe}
& \colhead{(\AA )}
& \colhead{(\AA )}
& \colhead{(\AA )}
& \colhead{(\AA )} 
}
\startdata
Q1004+1303 
& $0.24\pm0.03$ 
& $0.24\pm0.03$ 
& $0.40\pm0.05$ 
& $0.23\pm0.06$ \nl
Q0957+1317
& $0.21\pm0.05$
& $0.16\pm0.05$
& $0.40\pm0.10$
& $0.21\pm0.11$ \nl
Q1008+1319
& blended?
& $<0.41$
& $<1.18$
& $<1.37$ \nl
\tablenotetext{a}{All limits are $2\sigma(W)$}
\enddata
\end{deluxetable}


\clearpage
\begin{deluxetable}{lcc}
\tablenum{3}
\tablewidth{0pc}
\tablecaption{COLUMN DENSITY LIMITS TOWARD Q1004+1303}
\tablehead{
\colhead{Ion} & \colhead{$\log N$} & \colhead{$\log N$(H)} 
}
\startdata
Si~II	& $<13.6$	& $< 19.0$ \nl
Si~IV	& $<13.1$	& $< 18.5$ \nl
C~IV	& $<13.6$	& $< 18.0$ \nl
Mg~II   & $<12.0$	& $< 17.4$ \nl
\tablecomments{$\log N$(H) is the value deduced assuming that the particular
ion dominates the ionization fraction of the element. 
}
\enddata
\end{deluxetable}


\clearpage

\figcaption{Reproduction of a UK Schmidt plate showing the relative positions
of two of the three QSO/AGNs observed with \hst, 
Q1008+1319 (marked top left of the print) and
Q1004+1303 (marked to the right). 
Leo~I itself can be seen directly north of Regulus, the star which dominates
the bottom of the print.
For scale, the separation between Q1004+1303
and the center of Leo~I is 34.0 arcmins. NE is top left of this figure. 
\label{fig:poss}
}

\figcaption{Portions of the normalized G140L 
spectra of the 3 QSO/AGNs observed at 
the wavelength region
expected for Si~IV absorption from Leo~I. Absorption is seen from gas in 
our own 
Milky Way, but no absorption is detected from Leo~I at or near a 
heliocentric velocity
of 285~\kms . \label{fig:spec1}
}

\figcaption{Same as Figure~2, except the wavelength region 
covers that expected for C~IV
and Si~II absorption. 
For Q1004+1303,
complex absorption between 1530 and 1540~\AA\ arises from N~V absorption close
the emission redshift of the QSO. C~IV and Si~II
absorption are detected from our own Galaxy, but 
none is detected from Leo~I. \label{fig:spec2}
}

\figcaption{ Upper limits for the mean hydrogen density, $\rho$, and total
mass $M_{\rm{H}}$, in the halo of Leo~I for a spherical (solid lines) and
shell (dotted) distribution as a function of the total extent of the halo,
$R_s$. The figure assumes a limiting column density of $log N$(H)$=18$. 
\label{fig:spec3}
}


\setcounter{figure}{1}

\clearpage
\begin{figure}
\centerline{\psfig
{figure=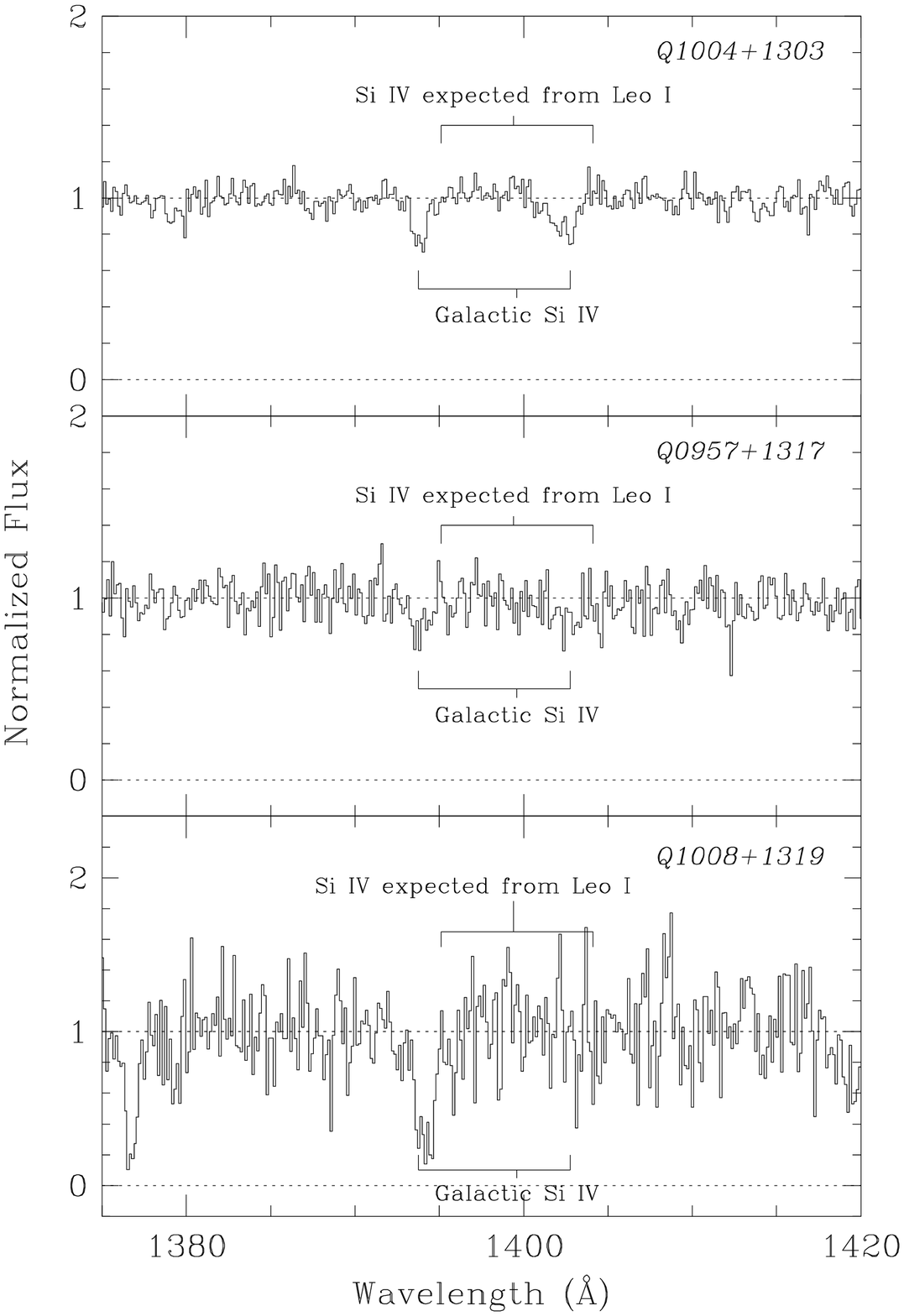,height=23cm}}
\caption{\label{fig:spec1}}
\end{figure}

\clearpage
\begin{figure}
\centerline{\psfig
{figure=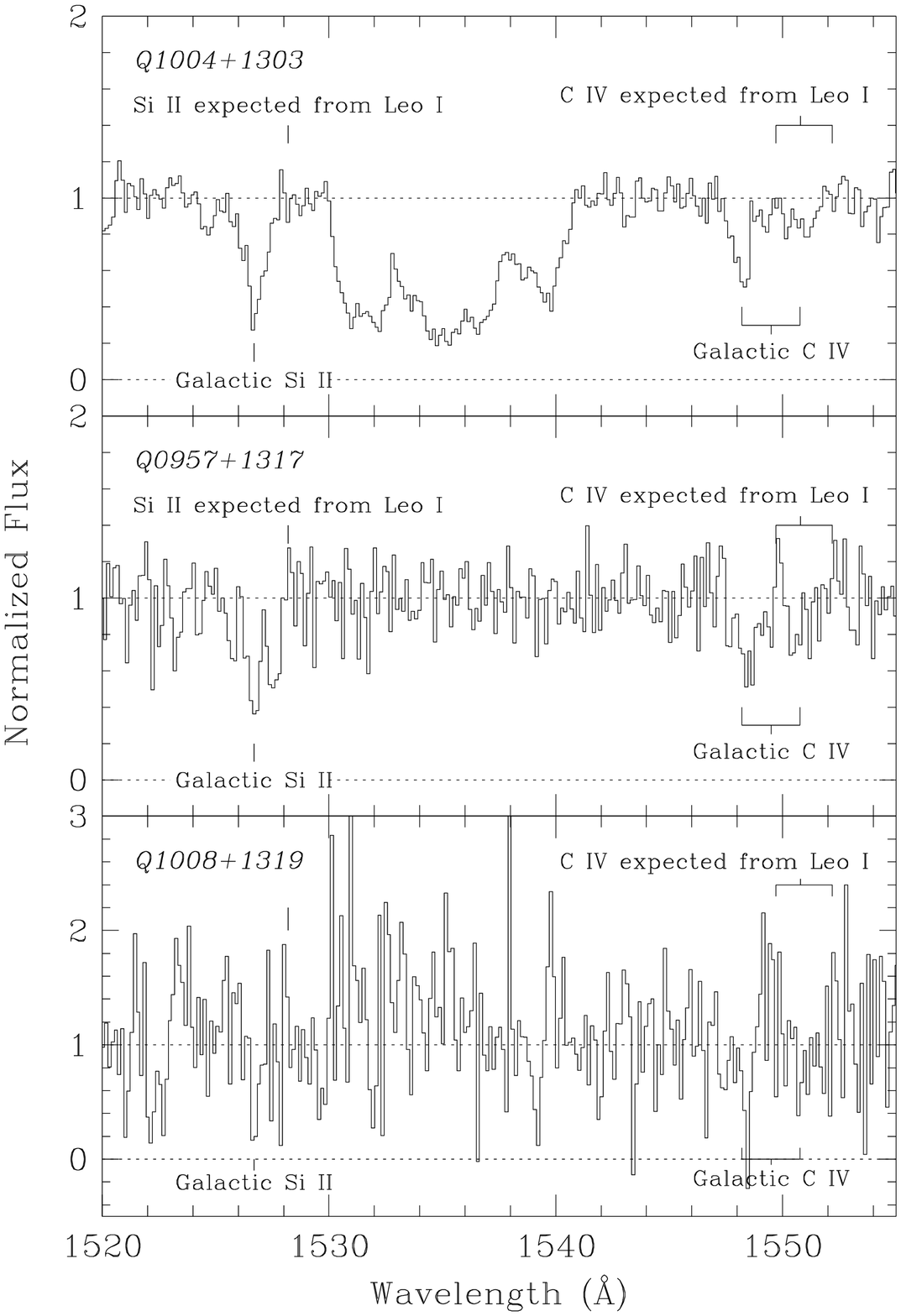,height=23cm}}
\caption{\label{fig:spec2}}
\end{figure}

\clearpage
\begin{figure}
\centerline{\psfig
{figure=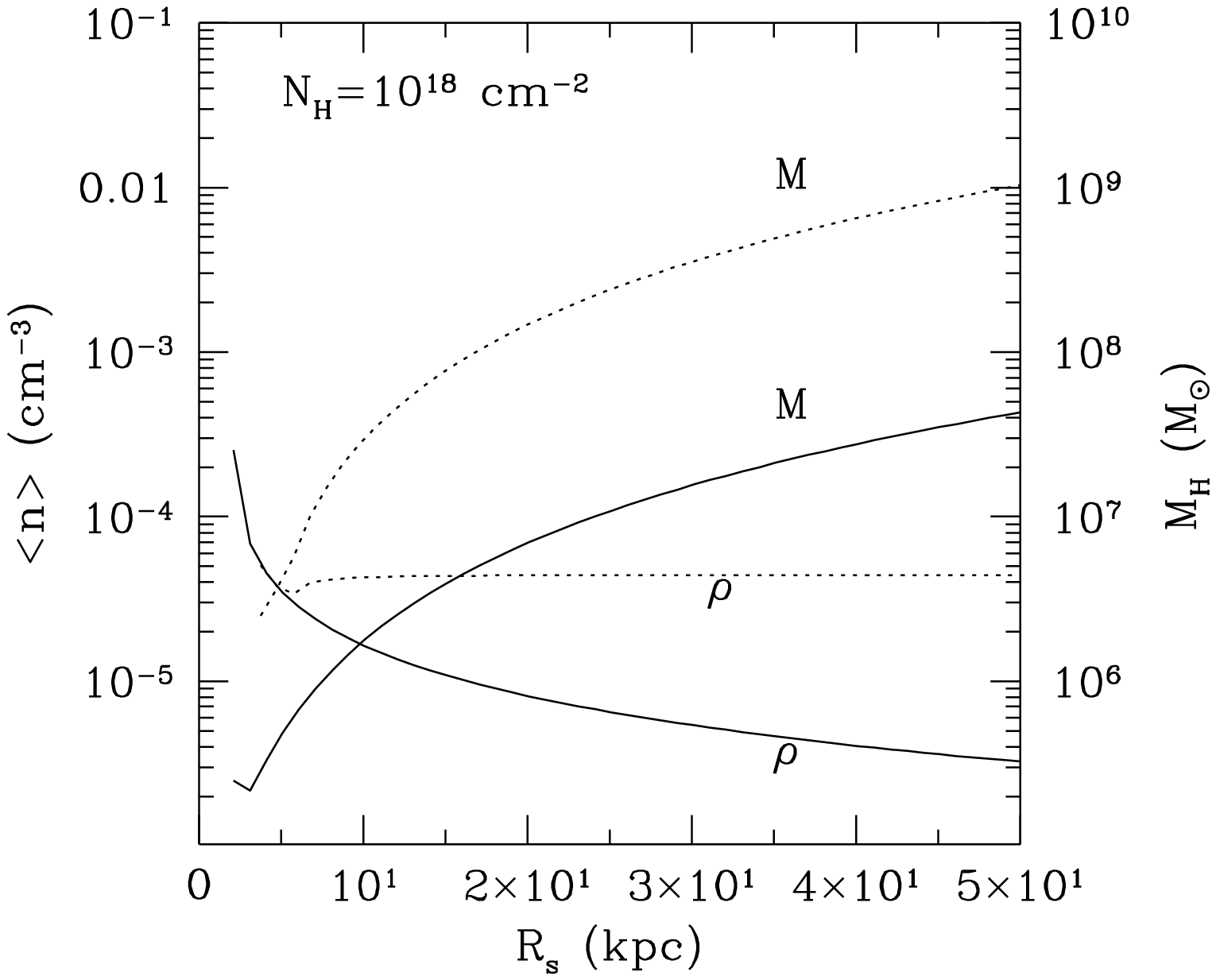,height=20cm}}
\caption{\label{fig:spec3}}
\end{figure}


\end{document}